\documentclass{ptephy}

\usepackage{boites}

\usepackage{array}
\usepackage{natbib}
\usepackage{bm}
\begin{document}

\title{Quark mean-field model for single and double $\Lambda$ and $\Xi$ hypernuclei}

\author{\name{\fname{J. N.} \surname{Hu}}{1,\ast}, \name{\fname{A.} \surname{Li}}{2,3,\ast}, \name{\fname{H.} \surname{Shen}}{4}, and \name{\fname{H.}
\surname{Toki}}{5}}

\address{\affil{1}{State Key Laboratory of Nuclear Physics and
Technology, School of Physics, Peking University, Beijing 100871,
China}
\affil{2}{Department of Astronomy and Institute of Theoretical
Physics and Astrophysics, Xiamen University, Xiamen 361005,
China}\affil{3}{State Key Laboratory of Theoretical Physics,
Institute of Theoretical Physics, Chinese Academy of Sciences,
Beijing, 100190}\affil{4}{epartment of Physics, Nankai University, Tianjin 300071, China} \affil{5}{Research Center for Nuclear Physics (RCNP), Osaka University, Ibaraki, Osaka 567-0047, Japan}
\email{liang@xmu.edu.cn}}

\begin{abstract}

The study of hypernuclei in the quark mean-field model are extended from single-$\Lambda$ hypernuclei to double-$\Lambda$ hypernuclei as well as $\Xi$ hypernuclei (double strangeness nuclei), with a broken SU(3) symmetry for the quark confinement potential. The strength of the potential for $u,d$ quarks is constrained by the properties of finite nuclei, while the one for $s$ quark is fixed by the single $\Lambda$-nucleus potential at the nuclear saturation density, which has a slightly different value. Compared to our previous work, we find that the introduction of such symmetry breaking improves efficiently the description of the single $\Lambda$ energies for a wide range of mass region, which demonstrates the importance of this effect on the hypernuclei study. Predictions for double-$\Lambda$ hypernuclei and $\Xi$ hypernuclei are then made, confronting with a few available experiential double-$\Lambda$ hypernuclei data and other model calculations. The consistency of our results with the available double-$\Lambda$ hypernucleus is found to be surprisingly good. Also, there is generally a bound state for $\Xi^0$ hypernuclei with both light and heavy core nuclei in our model.
\end{abstract}

\maketitle
\section{Introduction}

Theoretical studies of hypernuclei are continuously boosted by new and upgraded experimental facilities~\cite{LL1,LL2,LL3,LL4,LL5,Gal10,Bot12,Tam12}. It is generally believed that from them one could derive various features of the underlying hyperon interactions~\cite{Bod87,Usm04,Usm06,Hiy10,Hiy10s,Hiy12,Gal11,Li13}. They are related to the dense stellar matter studies~\cite{Li07,Hu13,Bur11} also, which mainly deal with strongly-interacting matter.

Lattice QCD calculations should be the ideal tool for investigating the hypernuclei structures since its input is the baryon-baryon interaction derived directly by the QCD theory. Indeed, the first calculation of hypernuclei with baryon number $A > 2$ has been performed recently~\cite{lqcd}, for $^4_{\Lambda}$He and $^4_{\Lambda\Lambda}$He. However, a detailed and precision structure description is still beyond its reach. Few-body calculations in the cluster or shell-model approaches are awaited for the studies of not-so-light (e.g., $A > 10$) hypernuclei. Significant progress in the auxiliary field diffusion Monte Carlo method~\cite{Lon13} has been done with the calculation of closed shell $\Lambda$ hypernuclei from $A = 5$ to $91$. For a more feasible way of the systematic study of both light and heavy hypernuclei, effective models are generally employed. Among them, many are for single-$\Lambda$ hypernuclei, for example, the quark mean-field (QMF) model~\cite{shen02}, the relativistic mean-field (RMF) approach~\cite{shen06,tokiptp,Xu12}, the Skyrme-Hartree-Fock (SHF) model~\cite{Li13,Sch13,Gul12}, the quark-meson coupling model~\cite{Gui08}, a relativistic point-coupling model~\cite{Tan12}, the quark mass density-dependent model~\cite{Wu13}. However, only a few theoretical calculations appear for double-$\Lambda$ hypernuclei, and hardly any for $\Xi$ hypernuclei. For example, $B_{\Lambda\Lambda}$ (two $\Lambda$ separation energies) prediction for $^6_{\Lambda\Lambda}$He to $^{13}_{\Lambda\Lambda}$B are available in the shell model~\cite{Gal11}. A self-consistent calculation of double-$\Lambda$ hypernuclei are presented in the RMF theory~\cite{shen06}. A few results for $\Delta B_{\Lambda\Lambda}$ (bond energies of double-$\Lambda$ hypernuclei) are shown lately in a SHF study~\cite{Sch13} with an effective $\Lambda$-nucleus interaction.  This $\Lambda$-nucleus interaction is microscopically derived from one of the most widely-used Brueckner-Hartree-Fock (BHF) calculations of isospin-asymmetric nuclear matter, using recently completed realistic Nijmegen hyperon potentials.

In the present work, based on our previous work~\cite{Hu13,shen00,shen02}, the study of hypernuclei in the QMF model are extended from single-$\Lambda$ hypernuclei to double-$\Lambda$ hypernuclei as well as $\Xi$ hypernuclei. Special efforts are devoted to introduce
effectively the SU(3) symmetry breaking. In the QMF model, the effect of gluon is introduced to the strength of a phenomenological confinement potential and constituent quark mass. This strength should also include pion cloud effect, which depends on quark flavors. The pion cloud effect is large for nucleons, but is small for hyperons. The SU(3) breaking effect can be effectively included in the strength of $s$ quark confinement potential. For this purpose, we assume a confining strength for the $s$ quark different from that for the $u,d$ quarks in the corresponding Dirac equations (under the influence of the meson mean fields).
The confining strength of $u,d$ quarks is constrained from properties of finite nuclei, and that of the $s$ quark by the well-established empirical value of $U^{(N)}_{\Lambda} \sim -30$ MeV. We study how this introduced breaking effect influence the single-$\Lambda$ properties. Also, based on the obtained fairly good reproduction of present single-$\Lambda$ hypernuclei data, we provide valuable predictions for double-$\Lambda$ hypernuclei and $\Xi$ hypernuclei. Comparisons are made with the available double-$\Lambda$ hypernuclei data~\cite{LL1,LL2,LL3,LL4,LL5} and results in other model calculations~\cite{shen06,Sch13}.

The paper is organized as follows. In Sec. II, we demonstrate our QMF model for the hypernuclei system. The numerical results and discussions are given in Sec. III. Finally, Sec. IV contains the main conclusions and future perspectives of this work.

\section{Formalism}

In the QMF model we shall take two steps to do the hypernuclei study. We first settle the problem of nucleons and $\Lambda,~\Xi$ hyperons in nuclear medium. Namely, these baryons as composites of three quarks are described in terms of the constituent quark model, in which the constituent quarks satisfy the Dirac equations with effective confinement potentials. Then in the second step, constructing the interaction between baryons through the meson fields, $\sigma$, $\omega$ and $\rho$, we study the multi-baryon hypernuclei system by knowing the individual baryon properties due to the presence of the mean fields from the first step.

The Dirac equations for constituent quarks can be written as:
\begin{eqnarray}\label{dirac}
 \left[ -i \vec \alpha \cdot \vec\nabla + \beta m_i^{*} +
 \beta \chi_c^i \right] q^i(r)=e_i^{*} q^i(r),
 \end{eqnarray}
where $i=q,s$ with the subscript $q$ denotes $u$ or $d$ quark. The
quark masses, $m_q=313\;\rm{MeV}$ and $m_s=490\;\rm{MeV}$, are
modified to $m_i^{*}=m_i+g^i_{\sigma}\sigma$ due to the presence of
the $\sigma$ mean field.
$e_i^{*}=e_i-g^i_{\omega}\omega-g^i_{\rho}\rho\tau^i_3$, with
$\sigma$, $\omega$, and $\rho$ being the mean fields at the middle
of the baryon. $e_i$ is the energy of the quark under the influence
of the $\sigma$, $\omega$, and $\rho$ mean fields. The confinement
potential is chosen to be a scalar-vector one as
$\chi^i_c=\frac{1}{2}k^ir^2 (1+\gamma^0)/2$. For the potential
strength, a previous study~\cite{shen02} of $\Lambda$ hypernuclei chose
$k^q=k^s=700 \;\rm{MeV/fm^2}$, applying the SU(3) symmetry. We here follow our recent work~\cite{Hu13} and adjust $k^s$ to reproduce properly hypernuclei experimental data. Because this symmetry breaking effect has been shown mainly manifested itself in the low density region of the stellar matter~\cite{Hu13}, we expect it would be essential to the hypernuclei structures, which is just around the nuclear saturation density.

By solving above Dirac equations, the change of the baryon mass $M_B^*$ as a function of the quark mass correction $\delta m_q=m_q-m_q^*$, can be studied~\cite{Hu13}. Then we can move to the second step. A single hypernucleus is treated as a system of many nucleons and one or two hyperons which
interact through exchange of $\sigma$, $\omega$, and $\rho$ mesons. The contribution of $\sigma$ meson is contained in $M_B^*$ and $\omega$ and $\rho$ mesons couple to the nucleons and hyperons ($\Lambda,~\Sigma,~\Xi$) with the coupling constants:
\begin{eqnarray}
&&g_{\omega N}=3g_\omega^q,\quad g_{\omega \Lambda}=cg_{\omega\Sigma}=2g_\omega^q,\quad
g_{\omega \Xi}=g_\omega^q \\
 & & g_{\rho N}=g_\rho^q,\quad g_{\rho \Lambda}=0,\quad g_{\rho \Sigma}=2g_\rho^q,\quad
 g_{\rho \Xi}=g_\rho^q
\end{eqnarray}
where the quark-meson coupling constants ($g^q_\sigma$, $g_\omega^q$, and $g_\rho^q$) and other basic parameters in the present model, have been determined by fitting the properties of nuclear matter, finite nuclei~\cite{shen00} and hypernuclei~\cite{Hu13}. Note here that in order to introduce the different behaviors between $\Lambda$ and $\Sigma$ hypernuclei, a factor $c$ is introduced before $g_{\omega\Sigma}$. From reproducing the single $\Sigma$ potential $U_{\Sigma}^{(N)}$ = 30 MeV at the nuclear saturation density~\cite{Sch00}, we choose $c = 0.775$. Namely, we take slightly larger $g_{\omega \Sigma}$ as compared to $g_{\omega \Lambda}$, to simulate the additional repulsion on the $\Sigma$-nucleon channel. When $c = 1$ it goes back to the standard choice usually employed~\cite{Hu13}. In other models~\cite{Miy09,Gui08,Oka87,Fuj01}, the one-gluon exchange potential is found to be very important for the description of $\Sigma$-hyperon potential (especially its repulsive nature).

From the QMF Lagrangian given
in Refs.~\cite{shen02,shen06}, we can write the equations of motion for baryons and mesons as
\begin{eqnarray}
&&\left[i\gamma_\mu\partial^\mu-M^*_N- g_{\omega B} \omega \gamma^0 -g_{\rho B} \rho
\tau_{3B}\gamma^0-e\gamma^0\frac{1+\tau_3}{2}A\right] \psi_B=0\\\nonumber
\label{eq:m1} &&(-\nabla^2+m^2_\sigma)\sigma+g_3 \sigma^3
= \sum_B \frac{\partial M_B^*}{\partial \sigma}\rho_S^B, \\\nonumber
\label{eq:m2} & & (-\nabla^2+m^2_\omega)\omega+c_3 \omega^3 = \sum_B g_{\omega B}
\rho_V^B, \\\nonumber
\label{eq:m3}& & (-\nabla^2+m^2_\rho)\rho = \sum_B g_{\rho B} I_{3B}\rho^B_V,\\\nonumber
&&-\nabla^2 A=e\rho_p
\end{eqnarray}
where $I_{3B}$ denotes the isospin projection
of baryon $B$. The $\rho^B_S$, $\rho^V_S$ and $\rho_p$ are the scalar, vector, and proton densities, respectively. The preceding coupling equations could be solved self-consistently for various hypernuclei systems. The hypernuclear system is restricted to the spherical case in this work and the pairing contribution for open shell nuclei is taken into account by using the BCS theory.

The center-of-mass correction is taken by the microscopic scheme~\cite{bender00},
\begin{eqnarray}
E_{\textrm{c.m.}}=\frac{\langle F|\bm{P}^2_{\textrm{c.m.}}|F\rangle}{2M_{\textrm{total}}}
\end{eqnarray}
where $M_{\textrm{total}}=\sum M_B=nM_\Lambda+(A-n)M_N$ is the total mass of the hypernuclei system including $n\Lambda$ hyperons, while $\bm{P}_{\textrm{c.m.}}=\sum \bm{P}_{\textrm{B}}$ is the total momentum operator. The expectation value of the square of the total momentum operator is calculated from the actual wave function of the ground state of the hypernuclei.

\section{Results}

The potential strength of strange quark, $k^s$, must be equal to the strength of $u,d$ quark, $k^q$, if the SU(3) symmetry is considered. However, the SU(3) symmetry is not strictly conserved in nuclear physics, e.g. the mass difference between nucleon and $\Lambda$ hyperon. Therefore, the strange potential strength, $k^s$, may differ from the $u,d$ quark case to take the effect of SU(3) symmetry breaking. Particularly, the potential strength should contain both the quark and pion cloud effects. The $k^q$ in QMF theory is usually determined by the ground state properties of finite nuclei~\cite{shen00}. Similarly, the magnitude of $k^s$ can be extracted from the properties of hypernuclei~\cite{Hu13}, such as $\Lambda$ hypernuclei, which is well known in the strangeness physics and have rich experimental data. Its single particle potential, $U^{(N)}_\Lambda$, is around $-30$ MeV, at the nuclear saturation density. With such a constraint, we can choose the strange potential strength in the QMF model as, $k^s=1000$ MeV/fm$^2$, which can generate the single $\Lambda$ potential as $U^{(N)}_\Lambda=-28.58$ MeV at saturation density, $\rho=0.145$ fm$^{-3}$. This strength of strange quark, $k^s$ is slightly different from the one chosen in nuclear matter case, $k^s=1100$ MeV/fm$^2$ in Ref.~\cite{Hu13}, that value would make the single $\Lambda$ hypernuclei overabound.

\begin{figure}[t!]
\centering
\includegraphics[width=0.7\textwidth]{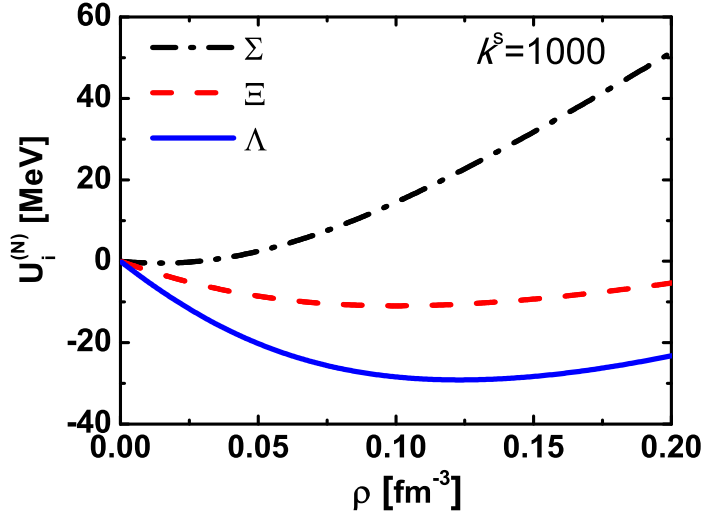}
\caption{(Color online) Single hyperon potentials $U^{(N)}_i$ as a function of density.}\label{fig1}
\end{figure}

In Fig.\ref{fig1}, the single hyperon ($\Lambda,\Sigma,\Xi$) potentials as a function of density are plotted with $k^s=1000$ MeV/fm$^2$.
The $\Xi$ potential is found to be a little bit weak about $10$ MeV around the nuclear saturation density, which is consistent with the experiments~\cite{E885}. As for the $\Sigma$ potential, it is always repulsive as caused by the use of slightly larger $\omega$ coupling strength. This is consistent with the experimental facts that no middle and heavier mass $\Sigma$-hypernuclei have been found.

\begin{figure}[t!]
\centering
\includegraphics[bb=0 115 250 225, width=1.0\textwidth]{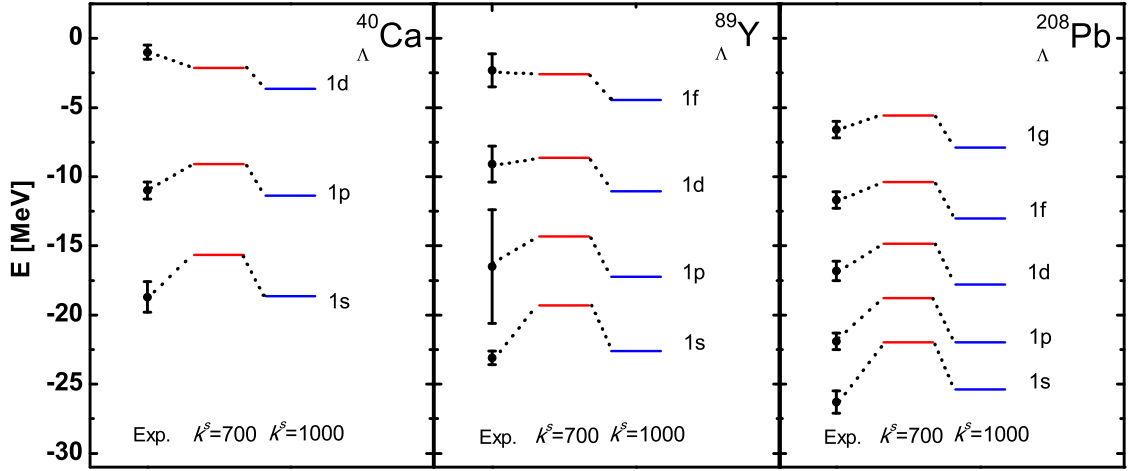}
\caption{(Color online) Energy levels of $\Lambda$ hyperon with different strange quark strengthes in the QMF model, compared with experimental data. }\label{fig2}
\end{figure}

With the new strange potential strength, $k^s=1000$ MeV/fm$^2$, we can calculate the single-$\Lambda$ hypernuclei system, whose properties are well known from experiments. The energy levels of $^{40}_\Lambda$Ca, $^{89}_\Lambda$Y and $^{208}_\Lambda$Pb are given in Fig.~\ref{fig2} and are compared with the experimental data and the theoretical results with $k^s=700$ MeV/fm$^2$ in the same method~\cite{shen02}. It is shown that the larger strange potential strength, the deeper $\Lambda$ binding state. Consequently the $\Lambda$ single particle energies are lower than those from the original QMF model, and more importantly, improved comparisons with the experiments are observed. Therefore, the effective SU(3) symmetry breaking effect is very important to the properties of hypernuclei in the QMF model.

Encouraged by the good agreements of single $\Lambda$ hypernuclei data in our SU(3)-symmetry-broken QMF model, we proceed to calculate also the $\Xi^0$ hypernuclei in the same framework, to serve as a reference for the future experiments. The single particle energies of $\Xi^0$ for $^{40}_{\Xi^0}$Ca, $^{89}_{\Xi^0}$Y and $^{208}_{\Xi^0}$Pb, are collected in the Table~\ref{tab2}, following the theoretical values and available experimental data of the corresponding $\Lambda$ hypernuclei. It can be seen that the binding energies of $\Xi^0$ are only $1/4\sim1/3$ of those of the $\Lambda$ hyperon. The deepest bound state of $\Xi^0$ exists in the $^{208}_{\Xi^0}$Pb, about $-8$ MeV.

\begin{table}[!htb]
\begin{center}
\caption{Single-particle energies (in MeV) for $^{40}_Y$Ca, $^{89}_Y$Y and $^{208}_Y$Pb in the QMF model and the corresponding experimental data. }\label{tab2}
\small
\renewcommand\arraystretch{1.5}
\begin{tabular}{p{0.8cm}|p{1.8cm}p{0.8cm}p{0.8cm}|p{1.8cm}p{0.8cm}p{0.8cm}|p{1.85cm}p{0.8cm}p{0.8cm}}
\hline\hline
             {}                 & {$^{40}_\Lambda$Ca (Exp.)} & ~ {$^{40}_\Lambda$Ca} & {$^{40}_{^0\Xi}$Ca}  &{$^{89}_\Lambda$Y (Exp.)} & ~{$^{89}_\Lambda$Y }&  {$^{89}_{^0\Xi}$Y }& {$^{208}_\Lambda$Pb (Exp.)}  &~{$^{208}_\Lambda$Pb}~&{$^{208}_{^0\Xi}$Pb}\\
\hline
             {$1s_{1/2}$}      &  {$-18.7\pm1.1$}            &  {$-18.64$}            &  {$-4.63$}            & {$-23.1\pm0.5$}          &{$-22.60$}           &  {$-6.31$}           & {$-26.3\pm0.8$}              &{$-25.37$}            &{$-7.78$}            \\
             {$1p_{3/2}$}       &  {             }            & {$-11.38$}            & {$-1.19$}            & {             }          & {$-17.24$}           &  {$-3.50$}           &{             }              &{$-21.97$}            &{$-5.82$}            \\
             {$1p_{1/2}$}       &  {$-11.0\pm0.6$}            &  {$-11.36$}            &  {$-1.18$}            & {$-16.5\pm4.1$}          & {$-17.22$}           &  {$-3.49$}           & {$-21.9\pm0.6$}              &{$-21.96$}            &{$-5.82$}            \\
             {$1d_{5/2}$}       &  {             }            & {$-3.65$}            &  {      }            &  {             }          & {$-11.05$}           &  {$-0.40$}           &{             }              &{$-17.80$}            &{$-3.46$}            \\
             {$1d_{3/2}$}       &  {$ -1.0\pm0.5$}            &  {$-3.62$}            &  {      }            &  {$ -9.1\pm1.3$}          & {$-11.03$}           &  {$-0.39$}           & {$-16.8\pm0.7$}              &{$-17.78$}            &{$-3.46$}            \\
             {$1f_{7/2}$}       &  {             }            & {      }            & {      }            &  {          }             &  {$-4.47$}            &  {     }            & {             }             &{$-13.05$}            &{$-0.85$}            \\
             {$1f_{5/2}$}       &  {             }            & {      }            &{      }            &  {$ -2.3\pm1.2$}          & {$-4.44$}            &  {     }            &{$-11.7\pm0.6$}             &{$-13.02$}            &{$-0.85$}            \\
             {$1g_{9/2}$}       &  {             }            & {      }            &  {      }            &  {             }          &  {     }            &  {     }            & {             }             &{$-7.90$}             &{     }             \\
             {$1g_{7/2}$}       &  {             }            & {      }            &  {      }            & {             }          &  {     }            & {     }            & {$ -6.6\pm0.6$}             &{$-7.86$}             &{     }             \\
\hline\hline
\end{tabular}
\end{center}
\end{table}

\begin{figure}[t!]
\centering
\includegraphics[width=0.7\textwidth]{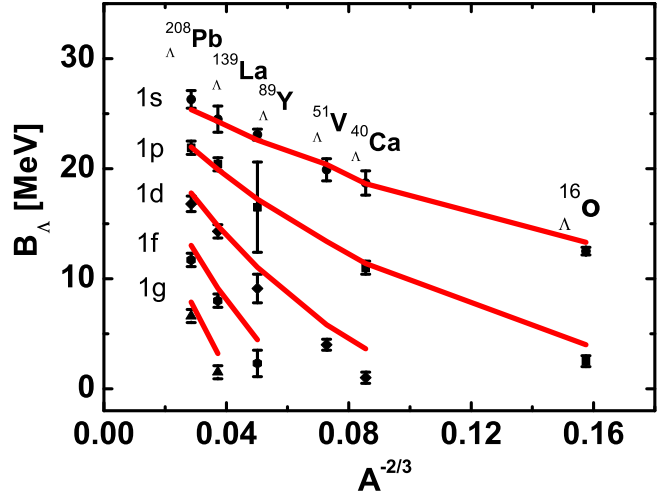}
\caption{(Color online) Systematic calculations of the binding energies of $\Lambda$ hypernuclei in the QMF model.}\label{fig3}
\end{figure}

Furthermore, we prepare in Fig.~\ref{fig3} the systematic calculation of the binding energies of $\Lambda$ hypernuclei, as a function of mass number $A$.
It is shown that the present QMF model (with $k^s=1000$MeV/fm$^2$) could reproduce fairly well the experimental values of $\Lambda$ binding energies in a wide range of mass number $A$, from $^{16}_\Lambda$O to $^{208}_\Lambda$Pb.

Then we move to study the properties of double-$\Lambda$ hypernuclei (shown in Table~\ref{dl}). For example, the $\Lambda\Lambda$ binding energy $B_{\Lambda\Lambda}$, which is obtained for the hypernucleus $^A_{\Lambda\Lambda}$Z by
\begin{eqnarray}
B_{\Lambda\Lambda}(^A_{\Lambda\Lambda}Z)=B(^A_{\Lambda\Lambda}Z)-B(^{A-2}Z),
\end{eqnarray}
and the $\Lambda\Lambda$ bond energy, $\Delta B_{\Lambda\Lambda}$, which is defined by
\begin{eqnarray}
\Delta B_{\Lambda\Lambda}(^A_{\Lambda\Lambda}Z)=B_{\Lambda\Lambda}(^A_{\Lambda\Lambda}Z)-2B_{\Lambda}(^{A-1}_{\Lambda}Z).
\end{eqnarray}
For comparison, the latest experimental data of light double-$\Lambda$ hypernuclei~\cite{nakazawa10,ahn13} are also included in the table. Those calculated results of heavy systems can be regarded as the predictions from our QMF model. Previous results from the RMF model are shown~\cite{shen06}, too. It is found that $B_{\Lambda\Lambda}$ increases with increasing mass number $A$, while $\Delta B_{\Lambda\Lambda}$ decreases at small mass numbers and increases at large mass numbers in the QMF model. For the light double-$\Lambda$ hypernuclei, our results of $B_{\Lambda\Lambda}=7.94$  MeV and $\Delta B_{\Lambda\Lambda}=0.75$ MeV for $^6_{\Lambda\Lambda}$He are very close the revised data of $Nagara$ event ($6.91\pm0.16$ and $0.67\pm0.17$, respectively). The calculations of $^{11}_{\Lambda\Lambda}$Be, $^{12}_{\Lambda\Lambda}$Be and $^{13}_{\Lambda\Lambda}$B are also consistent with experimental data. However, the value of $B_{\Lambda\Lambda}$ for $^{10}_{\Lambda\Lambda}$Be has a difference as large as 5 MeV with the corresponding experimental data. We notice that the experimental values between $^{10}_{\Lambda\Lambda}$Be and $^{11}_{\Lambda\Lambda}$Be differ relatively larger than those between $^{11}_{\Lambda\Lambda}$Be and $^{12}_{\Lambda\Lambda}$Be, and new experiments should be further needed for the values of $^{10}_{\Lambda\Lambda}$Be. In addition, the bond energies of double-$\Lambda$ hypernuclei are found to change sign with the increase of the mass number $A$ and saturate at $^{42}_{\Lambda\Lambda}$Ca.

\begin{figure}[t!]
\centering
\includegraphics[width=0.45\textwidth]{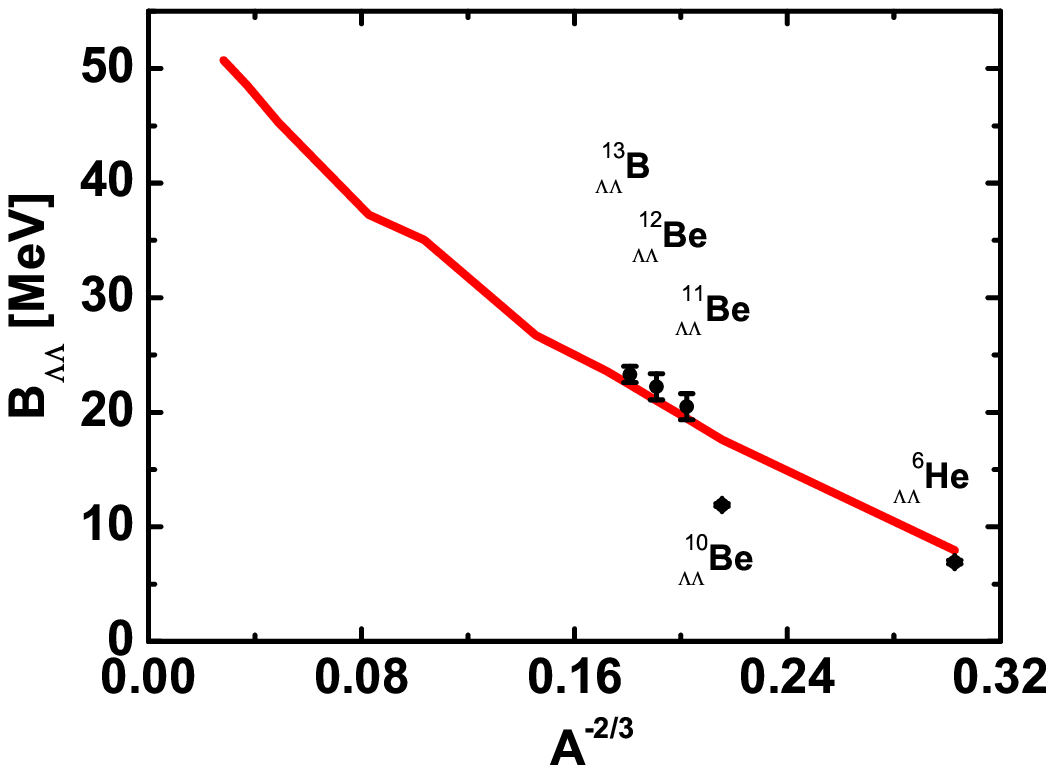}
\includegraphics[width=0.45\textwidth]{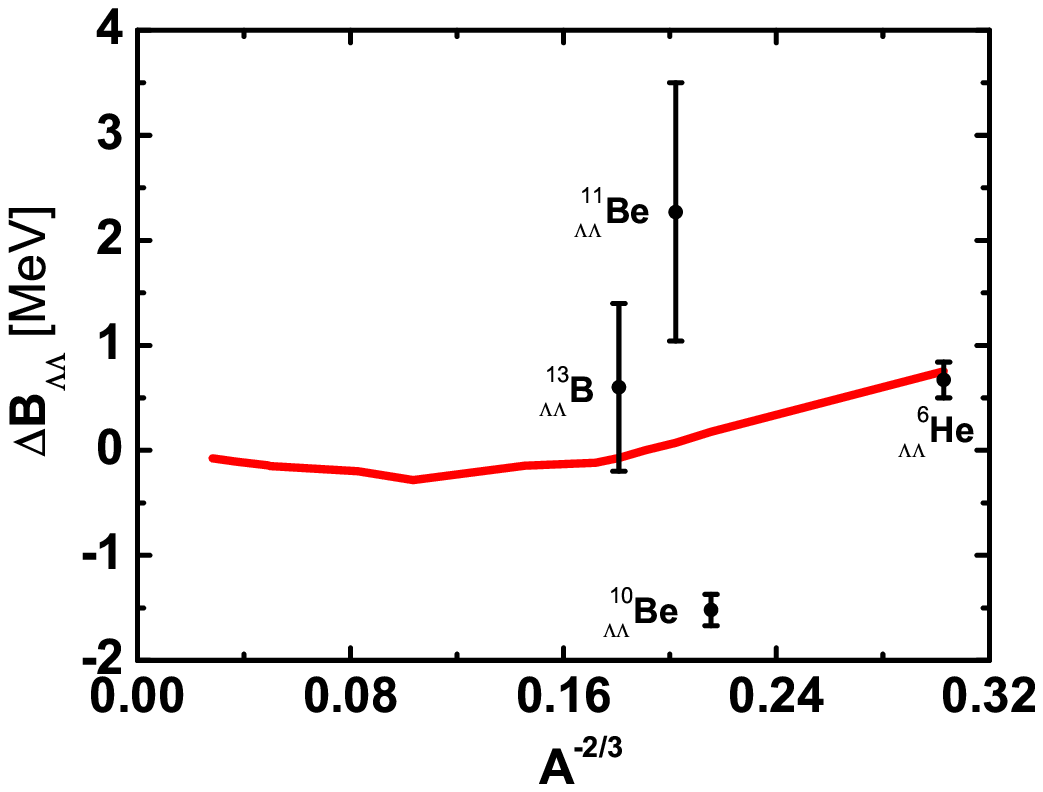}
\caption{(Color online) Systematic calculations of the binding energies and bond energies of $\Lambda\Lambda$ hypernuclei in QMF model.}\label{fig5}
\end{figure}

\begin{table}[!htb]
\begin{center}
\caption{$\Lambda\Lambda$ binding and bond energies, $B_{\Lambda\Lambda}$ and $\Delta B_{\Lambda\Lambda}$ (in MeV) respectively for double-$\Lambda$ hypernuclei for the QMF model, the RMF model and the available experimental data. }\label{dl}
\renewcommand\arraystretch{1.5}
\begin{tabular}{c c c c   c c c}
\hline\hline
            {$^A_{\Lambda\Lambda}$Z}    &\multicolumn{3}{c}{$B_{\Lambda\Lambda}$ } &\multicolumn{3}{c}{$\Delta B_{\Lambda\Lambda}$} \\
                     &~~  {Exp.}     &~~ {QMF} &~~ {RMF}
                       &~~  {Exp.}   &~~ {QMF} &~~ {RMF}  \\
\hline
             {$^6_{\Lambda\Lambda}$He}
              &~~  {$6.91\pm0.16$}   &~~  {7.94} &~~ {5.52}
              &~~  {$0.67\pm0.17$}     &~~  {$0.75$} &~~ {1.07}       \\
             {$^{10}_{\Lambda\Lambda}$Be}
             &~~  {$11.90\pm0.13$}    &~~ {17.61}  &~~ {16.34}
                 &~~  {$-1.52\pm0.15$}    &~~  {$0.18$}  &~~ {0.37}  \\
             {$^{11}_{\Lambda\Lambda}$Be}
              &~~  {$20.49\pm1.15$}   &~~ {19.46}   &~~ { }
                &~~  {$2.27\pm1.23$}   &~~  {$0.07$}  &~~ { }           \\
             {$^{12}_{\Lambda\Lambda}$Be}
             &~~  {$22.23\pm1.15$}  &~~ {21.00}     &~~ { }
             &~~ { } &~~  {$0.002$}    &~~ { }
                   \\
             {$^{13}_{\Lambda\Lambda}$B~}
              &~~  {$23.30\pm0.70$}    &~~ {22.41}     &~~ {22.14}
                 &~~  {$0.60\pm0.80$}  &~~  {$-0.07$}   &~~ {0.26}   \\
             {$^{42}_{\Lambda\Lambda}$Ca}
              &~~  {             }    &~~ {37.25}     &~~ {38.15}
                       &~~  {      }      &~~  {$-0.20$}  &~~ {0.04}     \\
             {$^{92}_{\Lambda\Lambda}$Zr}
               &~~  {             }      &~~ {45.24}  &~~ {47.11}
                     &~~  {      }   &~~  {$-0.14$}   &~~ {0.03} \\
             {$^{210}_{\Lambda\Lambda}$Pb}
                &~~  {             }    &~~ {50.73} &~~ {52.19}
                &~~  {      }    &~~  {$-0.08$}    &~~ {0.03}      \\
\hline\hline

\end{tabular}
\end{center}
\end{table}
Finally we presented in Fig.~\ref{fig5} the binding and bond energies of $\Lambda\Lambda$ hypernuclei as a function of the mass number $A$. When the mass number moves to infinite, the binding energies of double-$\Lambda$ hypernuclei approach $55$ MeV, i.e., twice of the corresponding single $\Lambda$ energy, while the bond energy goes to 0. The bond energies of double-$\Lambda$ have a saturation point at $^{42}_{\Lambda\Lambda}$Ca, which is different with the calculations with the RMF theory using TM1 and NL-SH parameter set~\cite{shen06}, but is consistent with the recent BHF calculations~\cite{Sch13}.

\section{Summary and future perspectives}

We have applied the QMF model to study the single-$\Lambda$, $\Xi^0$ and double-$\Lambda$ hypernuclei. The SU(3) symmetry breaking is considered in the quark level to be consistent with the experimental data of $\Lambda N$ potential at the nuclear saturation density, $U^{(N)}_\Lambda\sim-30$ MeV. This means that we have chosen the different potential strengthes for $u,d$ and $s$ quark at the quark mean field level. With such a strength for the strange quark, single $\Lambda$ and $\Xi^0$ potentials are both attractive at the saturation density, which implied the $\Xi$ hypernuclei may exist in the laboratory. We also increase slightly the $\Sigma$ coupling constant for $\omega$ exchange to achieve a positive value of $\Sigma$ hyperon potential at the nuclear saturation density of $30$ MeV.

We have calculated the single-$\Lambda$ hypernuclei within the present extended QMF model. The introduction of the SU(3) symmetry breaking results in a better reproduction of the experiment data than in the SU(3) symmetry case. In a rather wide mass range, the results of single-$\Lambda$ hypernuclei from the QMF model are quite consistent with the experimental values. The $\Xi^0$ hypernuclei of $^{40}_{\Xi^0}$Ca, $^{89}_{\Xi^0}$Y and $^{208}_{\Xi^0}$Pb were also calculated. There should be bound states for all these nuclei, and the binding energy was about 8 MeV for the deepest-bound in $^{208}_{\Xi^0}$Pb.

The double-$\Lambda$ hypernuclei are studied as well. The binding and bond energies of $^6_{\Lambda\Lambda}$He are very close to the revised data of the Nagara event. The other light double-$\Lambda$ hypernuclei are also reproduced very well for the binding energy. As to their bond energies, they at first decease with the increase of the mass number $A$, then increase with $A$ after the $^{42}_{\Lambda\Lambda}$ Ca, approaching 0 for more heavy nuclei. Although we have chosen the strength of strange quark confining potential from reproducing the single $\Lambda$-nucleon potential, there are not good justifications for the good agreements of the Nagara event ($^6_{\Lambda\Lambda}$He) and other double-$\Lambda$ hypernucleus. Further works are needed for this point, for example by tuning other parameters, or introducing hidden strange mesons in our model.

In summary, the hypernuclei system are discussed in the QMF model, and the available experimental data are reproduced rather well within our new QMF model. Many predications have been given for the heavy hypernuclei. With more facilities focusing on the hypernuclei study, the present QMF model will be applied for more aspects of them in the future, including possible further extensions and improvements.

\section*{Acknowledgment}
We would like to thank Dr. E. Hiyama for valuable discussions. The work is supported by the CPSC (Grant Nos. 2012M520100), and the Fundamental Research Funds for the Central Universities.

\bibliographystyle{plainnat}

\end{document}